# Ultrafast spatiotemporal chiroptical response of dielectric and plasmonic nanospheres

Ankit Kumar Singh[*,a], and Jer-Shing Huang [†,a,b,c,d]

a. Leibniz Institute of Photonic Technology, Albert-Einstein-Straße 9, 07745 Jena, Germany.
b. Abbe Center of Photonics and Institute of Physical Chemistry, Friedrich Schiller University Jena, Helmholtzweg 4, 07743 Jena, Germany.
c. Research Center for Applied Sciences, Academia Sinica, Taipei 11529, Taiwan.
d. Department of Electrophysics, National Yang Ming Chiao Tung University, Hsinchu 30010, Taiwan.

**Abstract**: We theoretically examine the spatiotemporal evolution of enhanced near-field optical chirality (OC) in both plasmonic and dielectric nanospheres when excited by ultrashort optical pulses. We demonstrate distinct spatiotemporal variations in near-field OC arising from the differing natures of plasmonic and dielectric resonators. The electric dipole resonant plasmonic nanosphere generates instantaneous near-field OC that relies on the interference between incident and scattered (induced) fields. Conversely, a resonant dielectric nanosphere sustains long-lasting OC even after the incident field diminishes due to the scattered field from resonant electric and magnetic dipole modes. We further demonstrate the control over the near-field OC using vector beams. Our work opens up opportunities for spatiotemporal control of nanostructure enhanced chiral-light matter interactions.

Chiral molecules can interact differently with chiral light, leading to chiroptical responses, like circular dichroism (CD). CD is the differential absorption of chiral light by chiral matter due to the electric-magnetic mixed polarizability in matter [1]. Therefore, CD is typically very weak and it depends simultaneously on the molecular structure and the properties of light. Enhancing CD by optical field engineering, in particular using rationally designed nanostructures to enhance optical chirality (OC) of the field, has drawn great attention in recent years. Recently, various dielectric and plasmonic nanostructures have been designed to provide an optical chirality enhancement (OCE) in the field around the nanostructures [2-6]. Since OC is directly proportional to the CD of chiral molecules [7], enhancing the OC by rationally designed nanostructures has been widely proposed to increase the sensitivity of chiroptical analysis [3-6,8-10].

Although efforts have been made to understand the structure-induced OC near nanostructures and their interaction with chiral molecules [11-15], the spatiotemporal evolution of the near-field OC around nanostructures excited by ultrashort laser pulses has not been investigated. Recently, femtosecond pulsed lasers are being used for probing the dynamical molecular properties with high temporal resolution [16,17]. In particular, several ultrafast studies were carried out to probe real-time chirality and chiral vibrational dynamics [18,19]. Very recently, a photoemission electron microscope has been used to investigate the time-dependent chiroptical response of plasmonic nanostructures [20]. However, how to link the intensity distribution obtained by photoemission electron microscope to the actual spatiotemporal evolution of the OC is still to be investigated. The understanding and control of the spatiotemporal evolution of OC near nanostructures may facilitate the investigation of ultrafast chiroptical light-matter interaction and the chiroptical detection based on temporal responses [21].

In this work, we theoretically study the spatiotemporal evolution of near-field OC around a dielectric (Si) and a plasmonic (Ag) nanosphere excited by a 2-fs broadband optical pulse. We use the Mie theory and finite-difference time-domain (FDTD) simulations to reveal distinct spatiotemporal OC responses and the different underlying mechanisms responsible for the OC generation of the two spheres. An electric dipole resonant plasmonic nanosphere requires the excitation/incident field to interfere with the induced/scattered field (SF) to provide an OC. Therefore, after the incident pulse vanishes, the OC around the resonant plasmonic nanoparticle drops rapidly. In contrast, a resonant dielectric nanosphere supports a high OC even after the excitation field vanishes due to the interference of the SF's of its electric and magnetic dipole mode. We further demonstrate the use of highly focused vector beams for exclusively excite the electric and magnetic modes with a controlled amplitude and phase difference to completely tune the OC distribution around the dielectric nanosphere. The presented study provides insight into the spatiotemporal evolution of OC and its tuning using highly focused vector beams.

For our study, we consider a nanosphere kept at the origin of a cartesian coordinate system excited with an x-polarized ideal plane wave source incident along the positive z-axis. Using Mie theory, the SF (electric field $\boldsymbol{E_s}$ and magnetic field $\boldsymbol{H_s}$) for incident x-polarized light can be written as [22],

$$\boldsymbol{E_s} = \sum_{n=1}^{\infty} E_n \left( i a_n \boldsymbol{N}_{e1n}^{(3)}(r,\theta,\phi) - b_n \boldsymbol{M}_{o1n}^{(3)}(r,\theta,\phi) \right) \quad (1)$$

$$\boldsymbol{H_s} = \frac{k}{\omega \mu} \sum_{n=1}^{\infty} E_n \left( i b_n \boldsymbol{N}_{o1n}^{(3)}(r,\theta,\phi) + a_n \boldsymbol{M}_{e1n}^{(3)}(r,\theta,\phi) \right)$$

Here, $E_n = i^n E_o (2n+1)/(n(n+1))$ is a term related to the incident field ($E_o$) and $n$ denotes the mode number. $k$ and $\omega$ represent the spatial and temporal frequency of the wave in free space, respectively. $\mu$ represents the permeability of the surrounding medium. The $a_n$ and $b_n$ are the Mie coefficients representing the electric and magnetic modes, respectively.

$N^{(3)}$ and $M^{(3)}$ denote the vector spherical harmonics with a dependence on the radial distance ($r$) from the origin, the angle from the x-axis ($\phi$) and the z-axis ($\theta$) [see supporting information (SI) for the functional form]. The equations give the value of the SF outside the spherical nanoparticle. The total field (TF, represented as $E$ and $H$) outside the nanosphere is a sum of the incident field ($E_i$ and $H_i$) and SF ($E_s$ and $H_s$). The symmetry of the problem can be used to calculate the fields for y-polarization, and hence for the circularly polarized electromagnetic fields (see SI).

The OC of an electromagnetic field can be obtained using the following equations [1,7],

$$C = \frac{\epsilon_o}{2} E \cdot \nabla \times E + \frac{1}{2\mu_o} B \cdot \nabla \times B \quad (2)$$
$$= -\frac{2\omega}{\epsilon_o} Imag(E^* \cdot B)$$
$$= -\frac{2}{\epsilon_o}\left(\frac{\partial E_R}{\partial t} \cdot B_R - \frac{\partial B_R}{\partial t} \cdot E_R\right)$$

Here, $B = \mu H$ and '$*$' is used to denote the complex conjugate. $E_R$ and $B_R$ represent the real part of electric field and magnetic flux density respectively. The above equation implies that for enhanced OC in the electromagnetic field, $E$ and $B$ should be parallel to each other and the phase difference between the parallel component should be close to $\pi/2$ [23].

It can be noted from equation 1 and equation 2 that if only one of the modes is excited by the incident light, for example, the electric dipole, $a_1$, in the case of resonant plasmonic nanoparticles, the OC of the SF obtained for linear polarized light (LPL) incidence is zero at all point in space (see detailed discussion in SI). However, this is not the same for the SF obtained from circularly polarized light (CPL) or the TF, regardless of excitation polarization. Exciting a plasmonic nanosphere with CPL, the OC of the near field comes from the incident field itself and the enhanced value of the OC is due to the field enhancement at the resonance. A more interesting case is exciting a plasmonic nanosphere (achiral object) with LPL (achiral light), OC enhancement can still be obtained in the near field (TF) of the nanosphere due to the interference between the incident and SF giving a four-lobe OCE pattern [24]. The OC is maximized close to the nanosphere's resonance frequency because the phase difference between the incident and the SF naturally reaches $\pi/2$ [23].

Fig. 1 shows the near-field optical response of a silver (radius 30 nm) and silicon (radius 100 nm) nanosphere in vacuum, for a LPL polarized at an angle of 45° with the x-axis, obtained using eq. 1 and 2. The optical constant of silver [25], and silicon [26], is taken from the experimental data. The dimension of the plasmonic nanosphere is chosen such that only electric dipolar mode ($a_1$) is excited at the wavelength ($\lambda$) of 365 nm and the contributions of magnetic dipole mode ($b_1$) and other higher order modes is minute, as shown in the inset of Fig. 1a. The extrema of OCE of SF and the interference of the electric and the magnetic dipole modes [Re($a_1 \cdot b_1^*$)] overlaps, showing a direct correlation between the two terms. However, a very small magnitude of the interference term (due to the small value of $b_1$) gives very small OC in the SF. Figs. 1b and 1c show the spatial distribution of the OCE calculated 1 nm above the surface of the nanosphere for SF and TF at the electric dipole resonant wavelength of 365 nm, respectively. The maxima of OCE of the TF shows a larger enhancement ($\approx 8$ times) when compared to SF at the same wavelength. The enhanced OC for the TF indicates the importance of the incident field presence for obtaining OC from the plasmonic scatterer.

Now, we look at the case of a dielectric silicon nanosphere. In this case, both electric and magnetic dipole modes can be excited and contribute concurrently to the SF of the nanosphere. The simultaneous excitation of the electric and magnetic modes increases the magnitude of the $a_1$ and $b_1$ mode interference (see Fig. 1d). The presence of the two dipolar modes also adds the otherwise absent contribution of the dominant radial electric and magnetic field component to the OC (see Eq 1, SI for details), allowing obtaining an increased OCE in the SF (see Fig. 1d). The maximum OCE is obtained close to the magnetic dipole resonance ($b_1$) and, showing a dependence on the mode interference term (Re($a_1 \cdot b_1^*$)). The interference of the magnetic and electric dipoles of the nanosphere leads to an OCE near the surface of nanosphere even without direct contribution of the incident field. Figs. 1e and 1f show the OCE enhancement distribution above the surface of nanosphere for SF and TF at a wavelength of 770 nm (magnetic dipole resonance), respectively. The SF alone already leads to high OCE, suggesting that interference of SF with the incident field does not play an important role, as in the case of plasmonic nanosphere.

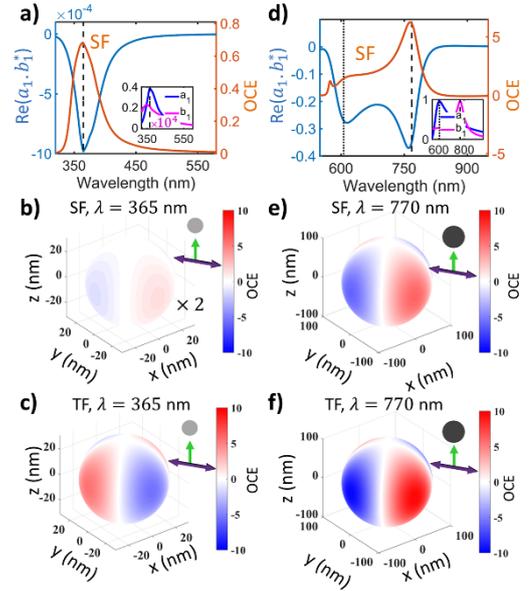

**Fig. 1**: The role of dipole mode interference on the OCE spectrum and the spatial distribution of (a-c) a silver nanosphere of radius 30 nm, and (d-f) a silicon nanosphere radius 100 nm, kept in vacuum for a linearly polarized light with polarization axis inclined at 45° with the x-axis. The effect of $a_1$ and $b_1$ mode interference (blue, left axis) on the calculated OCE spectrum (red, right axis) obtained at a point on y-axis 1 nm away from the nanoparticle's surface for silver nanosphere (a) and silicon (d) nanosphere. The inset shows the calculated spectrum of $a_1$ and $b_1$ mode. The calculated OCE distribution at a distance of 1 nm away from the surface of the nanosphere for silver nanosphere (b, c) and silicon nanosphere (d, f) considering only the SF (b, e) and the TF (c, f) electromagnetic field. The spatial OCE distribution is calculated at the wavelength mentioned on the top. The inset represents the polarization of light (blue double-headed arrow) and the direction of incidence (green arrow).

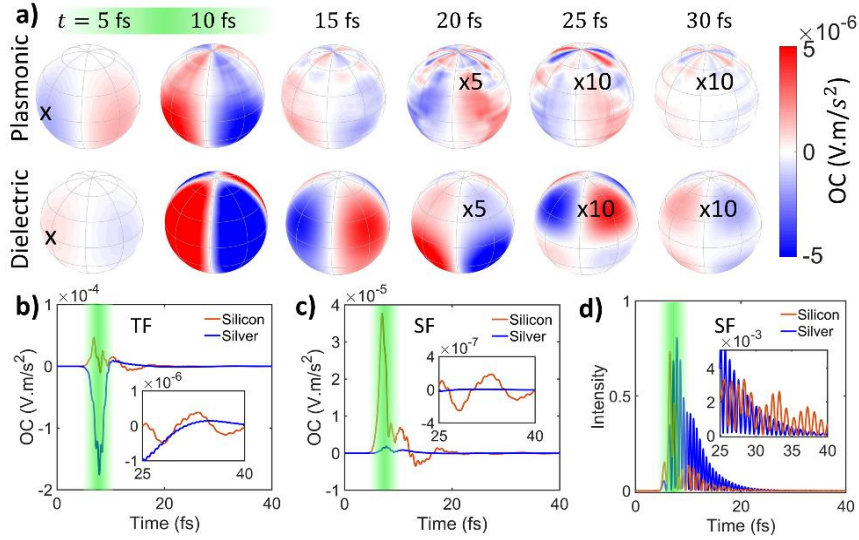

**Fig. 2**: The simulated spatiotemporal response of the electromagnetic field in the vicinity of the silicon and silver nanospheres obtained with FDTD simulation with 2 fs light pulse of LPL (polarization axis inclined at 45° with the x-axis) from a TFSF source. a) The spatial distribution of the OC pattern 1 nm away from the nanoparticle surface for the plasmonic (top row) and dielectric (bottom row) nanosphere at different instant of time (each 5 fs). The green strip is used to mark the light pulse intensity at the center of sphere. The OC values at longer time are scaled for better representation. The 'x' is used to mark the position of the angular coordinate of the point used for later discussion. The time-dependent OC of plasmonic and dielectric nanosphere obtained at the 'x' mark for (b) the TF (inside the TFSF source box), 1 nm from the surface of nanospheres, (c) the SF (outside the TFSF source box), 15 nm from the surface of nanospheres. d) The SF obtained outside the TFSF source at a distance of 15 nm from the surface of nanospheres. The values for longer time scales are zoomed in the inset of Figs. b-d.

Here, the size of the nanosphere is chosen to selectively observe the electric dipole resonances in the case of plasmonic and, electric and magnetic dipole in the case of dielectric for simplified discussion. The results of plasmonic nanosphere can be used to generalize to all achiral nanostructures where only the electric multipoles can be dominantly excited. In contrast, results from dielectric nanospheres are generally true for all achiral nanostructures where both electric and magnetic multipoles can be excited.

The presence of OCE in the SF of a nanoparticle without considering the incident field becomes very important for time domain chiral studies, for example, when a time-dependent excitation like an ultrafast pulsed laser is used to excite the system. In order to demonstrate it, we used finite difference time domain (FDTD) simulation to obtain the time domain response of the silicon and silver nanospheres kept in vacuum. We use a cuboidal total field scattered field (TFSF) source to illuminate the volume inside the TFSF domain with LPL. The TFSF box is designed such that the volume inside the box contains the contribution of the TF (SF and the incident field) while the volume outside the box only contains the SF. Further details of the simulation are discussed in SI.

Fig. 2a shows the spatiotemporal OC distribution of the plasmonic and dielectric nanosphere. It can be observed that the OC of the plasmonic sphere reduces significantly when the incident pulse vanishes from the vicinity of the particle. In contrast, due to the presence of magnetic dipole resonance, the dielectric nanosphere provides a larger OC even after the pulse vanishes. In order to confirm, we observe the OC for TF (see Fig. 2b) and the SF (see Fig. 2c) for the case of plasmonic and dielectric spheres. For the TF, although, the plasmonic particle shows a higher OC in the presence of pulse it goes to zero when the incident pulse vanishes, in contrast, the dielectric particle still shows a larger value even after the light pulse. Interestingly for the SF outside the TFSF source box (fields are not directly influenced by incident pulse), the plasmonic sphere shows negligible OC when compared to the dielectric particle even in the presence of incident pulse. This difference between the two nanoparticles is solely because of the presence of the magnetic mode in the dielectric nanosphere and not the lifetime of the modes or loss of the field enhancement in the SF. Fig. 2d shows the field intensity of the two cases, at the same point in the SF. It can be noted that the field obtained from plasmonic nanostructure is higher when compared to the dielectric case in the concerned timescale and the field for dielectric particle becomes larger only after 30 fs. This implies that the OCE obtained from the dielectric nanosphere after the incident pulse stems from the interference of the ED and MD sustained by the dielectric sphere.

The origin of OC purely from the modal interference of a nanoparticle opens up the possibility of manipulating the OC's spatiotemporal distribution by controlling the phase and amplitude relationship between the two modes. For simplicity, we explicitly control the phase and amplitude of electric and magnetic dipole mode in the plane wave scattering of the dielectric nanosphere. Fig.3 shows the OCE spectrum of the dielectric nanosphere at different phase difference (Φ) between electric and magnetic dipole and amplitude ratio (A) for the excitation of the magnetic dipole to the electric dipole mode. Controlling the phase and amplitude difference allows the manipulation of the OCE spectrum, even including totally flipping the handedness of the OCE spectrum.

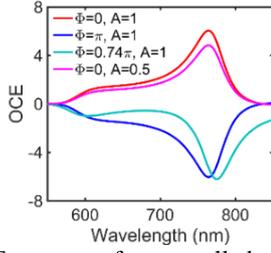

**Fig. 3**: The OCE spectrum for controlled phase difference and amplitude ratio of $a_1$ and $b_1$ modes in plane wave excitations.

Experimentally achieving this control is indeed feasible. It requires illumination schemes that can exclusively excite the electric modes and magnetic modes without affecting the vector field distribution. For example, the electric and magnetic modes can be exclusively excited by radial and azimuthal polarized vector beams [27], although it changes the field distribution. Nevertheless, it can still be used to experimentally demonstrate OCE manipulation. In the following we demonstrate such control in the SF of a dielectric nanosphere. The SF for highly focused radial ($E_s^{rad}/H_s^{rad}$) and azimuthal ($E_s^{azi}/H_s^{azi}$) polarized beam are given as [27,28],

$$E_s^{rad} = -\sum_{n=1}^{\infty} a_n \alpha_n N_{e0n}, \quad H_s^{rad} = \frac{ik}{\omega\mu}\sum_{n=1}^{\infty} a_n \alpha_n M_{e0n} \quad (3)$$

$$E_s^{azi} = -\sum_{n=1}^{\infty} b_n \alpha_n M_{e0n}, \quad H_s^{azi} = \frac{ik}{\omega\mu}\sum_{n=1}^{\infty} b_n \alpha_n N_{e0n}$$

Here,

$$\alpha_n = \frac{(kw_o)^2(2n+1)}{2\sinh\left(\frac{(kw_o)^2}{2}\right)} j_n\left(i\frac{(kw_o)^2}{2}\right)$$

is a factor related to the beam waist ($w_o$) of the incident beam and $j_n$ is the spherical Bessel function. The equation shows that the radial (azimuthal) polarized focused beam exclusively excite the electric (magnetic) modes of the nanosphere.

Fig.4 shows the OCE distribution for different phase delay ($\delta$) between the radial and azimuthal polarized beam. It shows that the handedness of the OCE distribution can be totally switched to the opposite by varying the phase difference from 0 to $\pi$ (Figs. 4a and b). With a special phase delay, it is possible to "turn off" the OC (Fig. 4c). The flipping of the handedness of OCE spatial distribution/ spectrum implies that a complete CD of an analyte could be obtained using controlled excitation. A similar OCE tuning can also be demonstrated by tuning the relative amplitude of the radial and azimuthal polarized beam. It must be noted that this amplitude and phase tuning of the two beams can be easily achieved in experiments by an attenuator and introducing extra optical path length in one of the beam paths, respectively. The possibility to manipulate the OCE distribution around small particles by shaping the excitation condition may also find applications in nanoscale circular dichroism measurement based on scanning probes. In addition, the use of nanoparticle enhanced chiral studies with vector beams can play vital role in the optimally enhanced CD measurements, that has shown advantages over the use of circular polarized light beams [29,30].

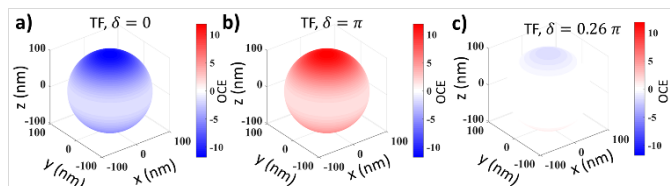

**Fig. 4**: The controlled tuning of OCE from the dielectric nanosphere by a phase delay of (a) 0, (b) $\pi$ and (c) $0.26\pi$ between the focused ($w_o = 350$ nm) radial and azimuthal beam excitation.

In conclusion, we have shown the evolution of OC in the temporal domain based on Mie theory and FDTD simulations. The study unravels the difference between the OCE enhancement obtained for nanostructures supporting only the electric multipole (plasmonic) and, both the electric and magnetic multipoles (dielectric). The presence of OC in the SF of a nanostructure can pave ways towards background-free (or without direct influence of incident light) CD measurement in the time domain (Fig. 2), and in controlling the spatial distribution in the frequency and time domain (Figs.3 and 4). The use of highly focused vector beams for obtaining OCE enhancement and complete switching of the handedness can also be used to directly measure CD of chiral system. It will be interesting to see the spatiotemporal response from higher order modes (with very large lifetime) of dielectric nanosphere, and other achiral and chiral nanostructures having the contribution of electric and magnetic resonances, or the add further control of the OCE distribution by using shaped ultrashort laser pulses [31]. The findings would stimulate further investigation into exploiting the modal properties of nanostructure to obtain a better and advanced system for OC studies, especially, in the temporal domain or for a pulsed excitation.

**Acknowledgement**: Financial support from the German Science Foundation (DFG) via Projects HU 2626/5-1 (No. 445415315), IRTG 2675 (No. 437527638), and SFB 1375/2 NOA-C1 (No.398816777) is gratefully acknowledged.

*AnkitKumar.Singh@leibniz-ipht.de
†Jer-Shing.Huang@leibniz-ipht.de

# Supporting Information
Ankit Kumar Singh[1,*], and Jer-Shing Huang [1,2,3,4,†]

[1]Leibniz Institute of Photonic Technology, Albert-Einstein-Str. 9, 07745 Jena, Germany

[2]Institute of Physical Chemistry and Abbe Center of Photonics, Friedrich Schiller University Jena, Helmholtzweg 4, 07743 Jena, Germany.

[3]Research Center for Applied Sciences, Academia Sinica, Taipei 11529, Taiwan.

[4]Department of Electrophysics, National Yang Ming Chiao Tung University, Hsinchu 30010, Taiwan.

**Section 1: Mie theory: electromagnetic scattering of plane waves**

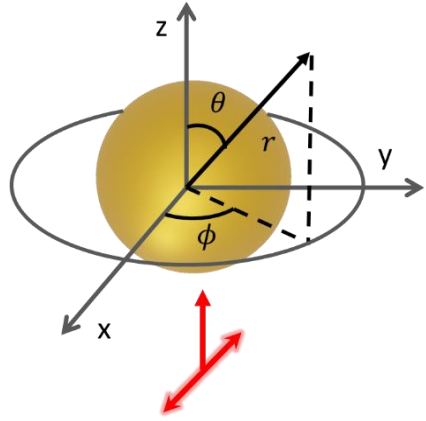

**Figure S1**: The schematic of scattering geometry. A linear polarized light is incident along the positive z-axis on the scatterer.

If a linearly polarized (x-polarized) plane wave is incident on an isotropic sphere kept in homogeneous environment, then using Mie theory, the expression for the scattered fields can be written as:

$$\mathbf{E}_s^x = \sum_{n=1}^{\infty} E_n \left( i a_n \mathbf{N}_{e1n}^{(3)}(r,\theta,\phi) - b_n \mathbf{M}_{o1n}^{(3)}(r,\theta,\phi) \right)$$
$$\mathbf{H}_s^x = \frac{k}{\omega\mu} \sum_{n=1}^{\infty} E_n \left( i b_n \mathbf{N}_{o1n}^{(3)}(r,\theta,\phi) + a_n \mathbf{M}_{e1n}^{(3)}(r,\theta,\phi) \right)$$

(S1)

where, $E_n = i^n E_o (2n+1)/(n(n+1))$, $E_0$ being the amplitude of the incident electric field. $k$, $\omega$, $\mu$ are respectively the wave number outside the sphere, frequency of the light and the permittivity of the surrounding medium; $a_n$ and $b_n$ are called the Mie co-efficient related to electric and magnetic multipoles respectively. **N** and **M** are the so-called vector spherical harmonics which satisfy the Helmholtz equation.

The vector spherical harmonics $\mathbf{N}^{(3)}$ and $\mathbf{M}^{(3)}$ can be decomposed along the three-unit vector ($\hat{r}$, $\hat{\theta}$, $\hat{\phi}$) in spherical polar coordinate system. Their forms are given below:

$$\mathbf{N}_{e1n}^{(3)} = \cos\phi\, n(n+1) \sin\theta\, \pi_n(\cos\theta) \frac{h_n^{(1)}(\rho)}{\rho} \hat{r} + \cos\phi\, \tau_n(\cos\theta) \frac{\left[\rho\, h_n^{(1)}(\rho)\right]'}{\rho} \hat{\theta}$$
$$- \sin\phi\, \pi_n(\cos\theta) \frac{\left[\rho\, h_n^{(1)}(\rho)\right]'}{\rho} \hat{\phi}$$

(S2)

$$\mathbf{N}_{e1n}^{(3)} = \sin\phi\, n(n+1) \sin\theta\, \pi_n(\cos\theta) \frac{h_n^{(1)}(\rho)}{\rho} \hat{r} + \sin\phi\, \tau_n(\cos\theta) \frac{[\rho h_n^{(1)}(\rho)]'}{\rho} \hat{\theta}$$
$$+ \cos\phi\, \pi_n(\cos\theta) \frac{[\rho h_n^{(1)}(\rho)]'}{\rho} \hat{\phi}$$

$$\begin{aligned}\mathbf{M}_{e1n}^{(3)} &= -\sin\phi\, \pi_n(\cos\theta)\, h_n^{(1)}(\rho) \hat{\theta} - \cos\phi\, \tau_n(\cos\theta) h_n^{(1)}(\rho) \hat{\phi} \\ \mathbf{M}_{o1n}^{(3)} &= \cos\phi\, \pi_n(\cos\theta) h_n^{(1)}(\rho) \hat{\theta} - \sin\phi\, \tau_n(\cos\theta) h_n^{(1)}(\rho) \hat{\phi}\end{aligned} \quad (S3)$$

where $\rho = kr$ and $h_n^{(1)}(\rho) = j_n(\rho) + i y_n(\rho)$ is the spherical Hankel function with $j_n(\rho)$ and $y_n(\rho)$ being the spherical Bessel functions.

For incident x-polarized light, we get the radial, polar and azimuthal components of the scattered electric field as:

$$\begin{aligned}E_{s\,r}^x &= \sum_{n=1}^{\infty} E_n\, i\, a_n \cos\phi\, n(n+1) \sin\theta\, \pi_n(\cos\theta) \frac{h_n^{(1)}(\rho)}{\rho} \\ E_{s\,\theta}^x &= \sum_{n=1}^{\infty} E_n \cos\phi \left[ i\, a_n\, \tau_n(\cos\theta) \frac{[\rho h_n^{(1)}(\rho)]'}{\rho} - b_n \pi_n(\cos\theta) h_n^{(1)}(\rho) \right] \\ E_{s\,\phi}^x &= \sum_{n=1}^{\infty} E_n \sin\phi \left[ -i\, a_n\, \pi_n(\cos\theta) \frac{[\rho h_n^{(1)}(\rho)]'}{\rho} + b_n \tau_n(\cos\theta) h_n^{(1)}(\rho) \right]\end{aligned} \quad (S4)$$

Similarly, the radial, polar and azimuthal components of the scattered magnetic field are:

$$\begin{aligned}H_{s\,r}^x &= \frac{k}{\omega\mu} \sum_{n=1}^{\infty} E_n\, i\, b_n \sin\phi\, n(n+1) \sin\theta\, \pi_n(\cos\theta) \frac{h_n^{(1)}(\rho)}{\rho} \\ H_{s\,\theta}^x &= \frac{k}{\omega\mu} \sum_{n=1}^{\infty} E_n \sin\phi \left( i\, b_n \tau_n(\cos\theta) \frac{[\rho h_n^{(1)}(\rho)]'}{\rho} - a_n\, \pi_n(\cos\theta)\, h_n^{(1)}(\rho) \right) \\ H_{s\,\phi}^x &= \frac{k}{\omega\mu} \sum_{n=1}^{\infty} E_n \cos\phi \left( i b_n \pi_n(\cos\theta) \frac{[\rho h_n^{(1)}(\rho)]'}{\rho} - a_n\, \tau_n(\cos\theta)\, h_n^{(1)}(\rho) \right)\end{aligned} \quad (S5)$$

The expression for the radial, polar and azimuthal components of the scattered fields for incident y-polarized light can be obtained by replacing $\phi$ by $(\phi - \pi/2)$ in the corresponding expressions for the incident x-polarized light.

For incident right circularly polarized (RCP) light, the scattered fields are obtained as $\mathbf{E}_s^{RCP} = \mathbf{E}_s^x + i \mathbf{E}_s^y$ and $\mathbf{H}_s^{RCP} = \mathbf{H}_s^x + i\, \mathbf{H}_s^y$. The expression of components of the fields are given below:

$$\begin{aligned}E_{s\,r}^{RCP} &= \sum_{n=1}^{\infty} e^{i\phi} E_n\, i\, a_n n(n+1) \sin\theta\, \pi_n(\cos\theta) \frac{h_n^{(1)}(\rho)}{\rho} \\ E_{s\,\theta}^{RCP} &= \sum_{n=1}^{\infty} E_n\, e^{i\phi} \left[ i\, a_n\, \tau_n(\cos\theta) \frac{[\rho h_n^{(1)}(\rho)]'}{\rho} - b_n \pi_n(\cos\theta) h_n^{(1)}(\rho) \right] \\ E_{s\,\phi}^{RCP} &= \sum_{n=1}^{\infty} E_n\, e^{i\phi} \left[ -a_n\, \pi_n(\cos\theta) \frac{[\rho h_n^{(1)}(\rho)]'}{\rho} - i\, b_n \pi_n(\cos\theta) h_n^{(1)}(\rho) \right]\end{aligned} \quad (S6)$$

$$H_r^{RCP} = \frac{k}{\omega\mu} \sum_{n=1}^{\infty} e^{i\phi}\, E_n\, b_n\, n(n+1) \sin\theta\, \pi_n(\cos\theta) \frac{h_n^{(1)}(\rho)}{\rho}$$
$$H_{s\,\theta}^{RCP} = \frac{k}{\omega\mu} \sum_{n=1}^{\infty} E_n\, e^{i\phi} \left( b_n \tau_n(\cos\theta) \frac{[\rho h_n^{(1)}(\rho)]'}{\rho} + i\, a_n\, \pi_n(\cos\theta)\, h_n^{(1)}(\rho) \right) \quad (S7)$$

$$H_{s\phi}^{RCP} = \frac{k}{\omega\mu} \sum_{n=1}^{\infty} E_n\, e^{i\phi} \left( ib_n \pi_n(\cos\theta) \frac{[\rho h_n^{(1)}(\rho)]'}{\rho} - a_n \tau_n(\cos\theta)\, h_n^{(1)}(\rho) \right)$$

For incident left circularly polarized (LCP) light, the scattered fields are obtained as $\mathbf{E}_s^{RCP} = \mathbf{E}_s^x - i\mathbf{E}_s^y$ and $\mathbf{H}_s^{RCP} = \mathbf{H}_s^x - i\,\mathbf{H}_s^y$.

## Section 2: Optical Chirality for the x-polarized and RCP beams for only electric and magnetic dipole modes.

The optical chirality ($= -2\omega\mu\, Imag(\mathbf{E}^*.\mathbf{H})/\epsilon_o$) of the field is the sum of three contribution each coming from radial, azimuthal and polar component. Assuming the electromagnetic scattering from the electric and magnetic dipole only (for simplicity), the three-individual contribution in the optical chirality of scattered field for x-polarized input is given as:

$$E_{s\,r}^{x*}. H_{s\,r}^{x} = \frac{k}{\omega\mu} \left(2\sin\theta\ \pi_n(\cos\theta)\right)^2 \left| E_1 \frac{h_1^{(1)}(\rho)}{\rho} \right|^2 a_1^* b_1$$

$$E_{s\,\theta}^{x*}. H_{s\,\theta}^{x} = \frac{k}{\omega\mu} \sin\phi \cos\phi\, |E_1|^2 \left( i\,\tau_n(\cos\theta)\pi_n(\cos\theta) \left( a_1^* a_1 \frac{[\rho h_n^{(1)}(\rho)]'^*}{\rho} h_n^{(1)} \right.\right.$$

$$\left.\left. - b_1^* b_1 \frac{[\rho h_n^{(1)}(\rho)]'}{\rho} h_n^{(1)*} \right) + a_1^* b_1 (\tau_n(\cos\theta))^2 \left| \frac{h_1^{(1)}(\rho)}{\rho} \right|^2 \right.$$

$$\left. + b_1^* a_1 (\pi_n(\cos\theta))^2 |h_n^{(1)}(\rho)|^2 \right) \qquad (S8)$$

$$E_{s\,\phi}^{x*}. H_{s\,\phi}^{x} = \frac{k}{\omega\mu} \sin\phi \cos\phi\, |E_1|^2 \left( -i\,\tau_n(\cos\theta)\pi_n(\cos\theta) \left( a_1^* a_1 \frac{[\rho h_n^{(1)}(\rho)]'^*}{\rho} h_n^{(1)} \right.\right.$$

$$\left.\left. - b_1^* b_1 \frac{[\rho h_n^{(1)}(\rho)]'}{\rho} h_n^{(1)*} \right) - a_1^* b_1 (\pi_n(\cos\theta))^2 \left| \frac{h_1^{(1)}(\rho)}{\rho} \right|^2 \right.$$

$$\left. - b_1^* a_1 (\tau_n(\cos\theta))^2 |h_n^{(1)}(\rho)|^2 \right)$$

From simple algebra, it becomes clear that the radial component ($E_{s\,r}^{x*}. H_{s\,r}^{x}$) will be non-zero if $a_n$ and $b_n$ are non-zero, the finding is also valid for circular polarization. Moreover, if either of the electric or magnetic multipoles are absent the angular contribution ($E_{s\,\theta}^{x*}. H_{s\,\theta}^{x} + E_{s\,\phi}^{x*}. H_{s\,\phi}^{x}$) also becomes zero for linear polarized light. This implies that for excitation of only electric or magnetic multipoles gives an identically zero optical chirality in the SF of linear polarized light and the dominant radial part of the field does not contribute in the chiral field generation.

## Section 3: FDTD simulation method

The FDTD simulation set up is shown in figure S2. A cuboidal total-field scattered-field (TFSF) source in the Lumerical FDTD is used to simulate the time domain optical response of the nanosphere. The size of the source is at least 15 nm larger than the diameter of nanosphere in all the directions. The polarization of light is kept at 45° with the x-axis in order to obtain the maximum optical chirality on the x-axis. The field inside the TFSF box was recorded using a 3-dimensional time monitor that is used to plot the OC snapshot in figure 2. A point time monitor was also placed outside the TFSF box (marked by 'x' mark) to record the scattered field only on the x-axis 15 nm from the surface of particle.

The boundary condition is set to PML to absorb all the scattered light. The plasmonic (dielectric) nanosphere is meshed with cubicle meshes of size 0.125(1) nm³. The FDTD simulation region is kept so that the PML layer is at a distance of at least 500 nm from the surface of the nanosphere. The simulation is terminated only after reaching the auto-shutoff value less than $10^{-5}$.

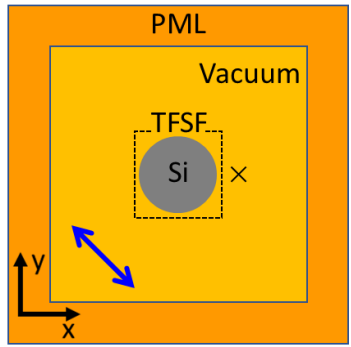

**Figure S2**: The FDTD simulation set-up. A TFSF source is used to illuminate the sphere kept in air and a physical match layer (PML) was used as the boundary condition. The cross-mark on the x-axis denotes a point monitor in the equatorial plane outside the TFSF domain used to obtain the electromagnetic field.